\documentclass[twocolumn,english,showpacs,preprintnumbers,amsmath,amssymb,floatfix]{revtex4-1}

\usepackage[T1]{fontenc}
\usepackage[latin9]{inputenc}
\usepackage{color}
\usepackage{array}
\usepackage{amstext}
\usepackage{graphicx}
\usepackage{esint}
\usepackage{rotating}
\usepackage[font=small,labelfont=bf]{caption}
\usepackage[colorlinks = true,
linkcolor = blue,
urlcolor  = blue,
citecolor = red,
anchorcolor = blue]{hyperref}

\makeatletter

\@ifundefined{textcolor}{}
{%
 \definecolor{BLACK}{gray}{0}
 \definecolor{WHITE}{gray}{1}
 \definecolor{RED}{rgb}{1,0,0}
 \definecolor{GREEN}{rgb}{0,1,0}
 \definecolor{BLUE}{rgb}{0,0,1}
 \definecolor{CYAN}{cmyk}{1,0,0,0}
 \definecolor{MAGENTA}{cmyk}{0,1,0,0}
 \definecolor{YELLOW}{cmyk}{0,0,1,0}
 }

\@ifundefined{definecolor}
 {\usepackage{color}}{}
\@ifundefined{definecolor}
 {\usepackage{color}}{}
\makeatother
\usepackage{babel}

\begin{document}


\title { QCD analysis of nucleon structure functions in deep-inelastic neutrino-nucleon scattering: Laplace transform and Jacobi polynomials approach  }

\author{ S. Mohammad Moosavi Nejad$^{a,c}$}
\email{Mmoosavi@yazd.ac.ir}

\author { Hamzeh Khanpour$^{b,c}$ }
\email{Hamzeh.Khanpour@mail.ipm.ir}

\author {S. Atashbar Tehrani$^{d}$}
\email{Atashbar@ipm.ir}

\author { Mahdi Mahdavi$^{a}$ }
\email{Mahdi.Mahdavi.j@gmail.com}

\affiliation {
$^{(a)}$Physics Department, Yazd University, P.O.Box 89195-741, Yazd, Iran          \\
$^{(b)}$Department of Physics, University of Science and Technology of Mazandaran, P.O.Box 48518-78195, Behshahr, Iran         \\
$^{(c)}$School of Particles and Accelerators, Institute for Research in Fundamental Sciences (IPM), P.O.Box 19395-5531, Tehran, Iran        \\
$^{(d)}$Independent researcher, P.O.Box 1149-8834413, Tehran, Iran   }

\date{\today}

%
%
\begin{abstract}\label{abstract}
We present a detailed QCD analysis of nucleon structure functions $xF_3 (x, Q^2)$, based on Laplace transforms and Jacobi polynomials approach.
The analysis corresponds to the next-to-leading order and next-to-next-to-leading order approximation of perturbative QCD. The Laplace transform technique, as an exact analytical solution, is used for the solution of nonsinglet Dokshitzer-Gribov-Lipatov-Altarelli-Parisi evolution equations at low- and large-$x$ values. The extracted results are used as input to obtain the $x$ and Q$^2$ evolution of $xF_3(x, Q^2)$ structure
functions using the Jacobi polynomials approach. In our work, the values of the typical QCD scale $\Lambda_{\overline{\rm MS}}^{(n_f)}$ and the strong coupling constant $\alpha_s(M_Z^2)$ are determined for four quark flavors ($n_f=4$) as well. A careful estimation of the uncertainties shall be performed using the Hessian method for the valence-quark distributions, originating from the experimental errors.
We compare our valence-quark parton distribution functions sets with those of other collaborations; in particular with the {\tt BBG}, {\tt CT14}, {\tt MMHT14} and {\tt NNPDF} sets, which are contemporary with the present analysis. The obtained results from the analysis are in good agreement with those from the literature.
\end{abstract}


\maketitle


%
%
\section{Introduction}\label{Introduction}

A unique view on the inner structure of the nucleons, through charged current interactions can be provided via high-energy neutrino-nucleon scattering. The measurements of neutrino-nucleon interactions can probe the quark-flavour structure of nucleons independently of the charged lepton scattering.
In fact, deep inelastic neutrino scattering allows one to map separately the parton distribution function for quarks and antiquarks in the  nucleon, while one can not determine the individual parton distribution functions (PDFs) from charged lepton scattering experiments alone.

The present data on the neutrino scattering has not yet reached the level of precision as is available  for the neutral-current deep-inelastic scattering (DIS).
However, the neutrino factories planned for the future will provide more precise data for deep-inelastic charged-current neutrino-nucleon scattering~\cite{Kaplan:2014xda,Bonesini:2016ncr,Geer:2010zz,Geer:2009zz,Banerjee:2015gca}. The data from these neutrino scattering experiments are crucial inputs for the global fits of  the parton distribution functions, which are essential for any calculation in high energy physics.

The nonsinglet structure functions $xF_3(x, Q^2)$, measured in CCFR~\cite{Seligman:1997mc} and NuTeV~\cite{Tzanov:2005kr} experiments at Fermilab and the recent neutrino oscillation search by the CHORUS collaboration~\cite{Onengut:2005kv} at CERN, have provided a precise experimental source to determine the valence-quark
distributions $x u_v (x, Q^2)$ and $x d_v (x, Q^2)$ of the nucleon.
Due to the absence of gluonic effects in the QCD evolution of the strong coupling constant ~\cite{Khorramian:2009xz,Blumlein:2004ip}, the flavor nonsinglet distributions allow for clear measurements of $\alpha_s$.
Moreover, it is expected that the nonsinglet data from charged-current interactions at the electron-proton collisions at the HERA collider and, in future, at  the Large Hadron Electron Collider (LHeC) at CERN~\cite{AbelleiraFernandez:2012cc,AbelleiraFernandez:2012ni} and Electron Ion Collider (EIC)~\cite{Aschenauer:2016our,Deshpande:2016goi,Accardi:2012qut}, will allow us to determine the valence-quark distributions $x u_v (x, Q^2)$ and $x d_v (x, Q^2)$ of the nucleon and strong coupling constant, with unprecedented precision.

In our earlier work~\cite{Khanpour:2016uxh}, we have made an extensive comparative study on the applicability of the Laplace transform technique to obtain the analytical solution for the proton structure function, $F_2^p (x, Q^2)$.
In the following manuscript, we present a detailed QCD analysis of charged-current structure functions $xF_3(x, Q^2)$ measured in neutrino-nucleon scattering at CCFR, NuTeV and CHORUS experiments.
The main new ingredient of the present analysis comes from the recent results on the analytical
calculations of next-to-next-to-leading order (NNLO) correction using the Laplace transform technique.
The Jacobi polynomials are also used to obtain the $x$ and Q$^2$ evolution of the structure functions $xF_3(x, Q^2)$.
In this auxiliary technique, the Jacobi polynomial approach is employed to achieve the $xF_3(x, Q^2)$
structure function in the ($x$, Q$^2$) plane while we used the
moments of valence-quarks densities in $s$-(Laplace) space.
Following these methods, using the Laplace transformation and Jacobi polynomial approach, we indicate
that these methods work well in which we were able to
extract the valence-quarks distribution functions form the global QCD analysis of neutrino-nucleon scattering data.
We will also perform a careful estimation of the uncertainties using the Hessian method for the valence-quark
distributions originating from the experimental errors.

The rest of the present paper is organized as follows:
In Sec.~\ref{framework}, we present the theoretical framework including the nonsinglet solution of Dokshitzer-Gribov-Lipatov-Altarelli-Parisi (DGLAP) evolution equations in Laplace space at the NNLO approximation and the Jacobi polynomial approach. Sec.~\ref{global-PPDFs} contains the method of the analysis, data selection, minimization and error calculations.
The results of our analysis is presented in Sec.~\ref{Sec:Results}. We give our summary and conclusions in Sec.~\ref{Conclusions}.
Appendix {\bf A} deals with a technical detail including the Laplace transform of the splitting functions for the nonsinglet sectors and the Wilson coefficient functions.

%
%
\section{ Theoretical framework }\label{framework}

In this manuscript, a detailed QCD analysis is performed using the repeated Laplace transform to find an analytical solution of the (DGLAP) evolution equations~\cite{Gribov:1972ri,Lipatov:1974qm,Dokshitzer:1977sg,Altarelli:1977zs} for the nonsinglet sector at the NNLO approximation. This newly-developed Laplace transform technique will be explained in detail in the next section.

%
%
\subsection{ Nonsinglet solution in Laplace space at the NNLO approximation } \label{Laplace}

In this work we, specifically, concentrate on the charged-current structure functions $xF_3(x, Q^2)$ at the next-to-leading order (NLO) and NNLO approximations in the perturbative QCD framework.
In recent years, using the Laplace transform technique, some analytical solutions of the DGLAP evolution equations
have been reported~\cite{Block:2010du,Block:2011xb,Block:2010fk,Block:2009en,Block:2010ti,Zarei:2015jvh,Boroun:2015cta,Boroun:2014dka} which reached considerable phenomenological success. There is also  some progress to extract the analytical solutions of the proton spin-independent structure function $F_2^p (x, Q^2)$~\cite{Khanpour:2016uxh} and the spin-dependent one $x g_1^p (x, Q^2)$~\cite{AtashbarTehrani:2013qea}, using the Laplace transform technique.

The DGLAP evolution equations for the flavor-nonsinglet (NS) sector have the following standard forms,
\begin{eqnarray}\label{eq:nonsinglet-integrate}
\frac{4 \pi}{\alpha_s(Q^2)}  \frac{\partial F_{\text{NS}}}{\partial \ln Q^2} (x,Q^2)=\int_x^1 \frac{dz}{z} F_{\text{NS}}(z, Q^2)\frac{x}{z}p_{qq}^{\text{NS}} \left(\frac{x}{z} \right) \,,  \nonumber \\
\end{eqnarray}
where $\alpha_s(Q^2)$ is the running coupling constant and  $P_{ij}^{\text{LO}}(\alpha_s(Q^2))$, $P_{ij}^{\text{NLO}}(\alpha_s(Q^2))$ and $P_{ij}^{\text{NNLO}}(\alpha_s(Q^2))$ are the nonsinglet Altarelli-Parisi splitting kernels at one, two and three loops corrections which satisfy the following expansion

  \begin{eqnarray}\label{eq:nonsinglet-Spliting}
  	P_{ij} (\alpha_s(Q^2)) & = & P_{ij}^{\text{LO}} (x) + \frac{\alpha_s(Q^2)}{2 \pi} P_{ij}^{\text{NLO}} (x)   \nonumber \\
  	& + &  (\frac{\alpha_s(Q^2)}{2 \pi})^2 P_{ij}^{\text{NNLO} } (x) \,.
  \end{eqnarray}
The coupled DGLAP evolution equations at the leading order (LO), NLO and NNLO contributions for the nonsinglet sector $F_{\rm NS}(x,Q^2)$ using convolution symbol $\otimes$ can be written as,
\begin{eqnarray}\label{eq:nonsinglet-DGLAP}
&&\frac{4 \pi}{\alpha_s(Q^2)}  \frac{\partial F_{\text{ NS}}}{\partial \ln Q^2} (x,Q^2)  = F_{\text{ NS}} \otimes \Big(p_{qq}^{\text{LO,NS}}  \nonumber  \\
&&+ \frac{\alpha_s(Q^2)}{4\pi} p_{qq}^{\text{NLO,NS}}
+ (\frac{\alpha_s(Q^2)}{4\pi})^2 p_{qq}^{\text{NNLO,NS}} \Big)(x,Q^2) \,.
\end{eqnarray}

We are now in a position to briefly summarize the method of extracting the valence-quark distribution functions via an analytical solution of the DGLAP evolution equations using the Laplace transform technique.

Considering the variable definitions $\nu \equiv \ln(1/x)$ and $w \equiv \ln(1/z)$, one can rewrite the evolution  equations (\ref{eq:nonsinglet-DGLAP}) in terms of the convolution
integrals and the new variables $\nu$ and $\tau$. Therefore, one obtains a simple solution as
 \begin{eqnarray}\label{eq:nonsinglet}
 &&	\frac{\partial \hat{F}_{\text{NS}}}{\partial\tau}(\nu,\tau)   =  \int_0^\nu   \Big(p_{qq}^{\text{LO, NS}}(\nu-w)   \nonumber \\
 	&&+ \frac{\alpha_s(\tau)}{4\pi}  p_{qq}^{\text{NLO,NS}}(\nu-w) + (\frac{\alpha_s(\tau)}{4\pi})^2  p_{qq}^{\text{NNLO,NS}}(\nu - w) \Big)  \nonumber \\
 	&&\times \hat{F}_{\text{NS}}(w, \tau)e^{-(\nu - w)} \, dw \,.
 \end{eqnarray}
The $e^{-(\nu-w)}$ factor in above equation comes from the $\frac{x}{z}$ term of Eq.~(\ref{eq:nonsinglet-integrate}) considering variable definitions $\nu \equiv \ln(1/x)$ and $w \equiv \ln(1/z)$.
The $Q^2$ dependence of evolution equations presented in Eq.~(\ref{eq:nonsinglet}), is expressed entirely thorough the variable $\tau$ as
$\tau (Q^2, Q_0^2)  \equiv \frac{1}{4 \pi} \int_{Q_0^2}^{Q^2} \alpha_s (Q'^2)  d\ln Q'^2$.
Defining the Laplace transforms $f_{NS}(s,\tau) \equiv  {\cal L} [ \hat F_{NS}(\nu,\tau); s]$ and considering the fact that the Laplace transform of the convolution factors is simply the ordinary product of the Laplace transform of the factors~\cite{Block:2010du,Block:2011xb}, the Laplace transform of Eq.~(\ref{eq:nonsinglet}) converts to the ordinary first order differential equations in Laplace space $s$ with respect to $\tau$ variable. Consequently, by working in the Laplace space  $s$, we can obtain the first order differential equations with respect to the variable $\tau$ for the nonsinglet distributions $f_{\rm NS}(s, \tau)$ as
 \begin{eqnarray}\label{eq:nonsinglet-laplace-space}
 	\frac{\partial f_{\text{NS}}}{\partial \tau}(s, \tau) && = \left (\Phi_{\text{NS}}^{\text{LO}} + \frac{\alpha_s(\tau)}{4\pi}\Phi_{\rm NS,qq}^{\text{NLO}}+ (\frac{\alpha_s(\tau)}{4\pi})^2\Phi_{\text{NS,qq}}^{\text{NNLO}} \right )   \nonumber  \\
 	&&\times f_{\text{ NS}}(s,\tau)\,.
 \end{eqnarray}
A very simplified solution of the above equation, reads
 \begin{eqnarray}\label{eq:solve-nonsinglet}
f_{\text{ NS}}(s, \tau) = e^{\tau \, \Phi_{\text{NS}}(s)} \, f^0_{\text{NS}}(s)\,,
 \end{eqnarray}
 where the $\Phi_{\text{ NS}}(s)$ contains contributions of the $s$-space splitting functions up to the NNLO approximation. These splitting functions can be calculated from $x$-space results, presented in  Refs.~\cite{Curci:1980uw,Moch:2004pa}. The result reads
 \begin{eqnarray}\label{eq:fi-nonsinglet}
 	\Phi_{\text{NS}}(s) \equiv \Phi_{\text{NS}}^{\text{LO}}(s) + \frac{\tau_2}{\tau}\Phi_{\text{NS,qq}}^{\text{NLO}}(s)+\frac{\tau_3}{\tau}\Phi_{\text{NS,qq}}^{\text{NNLO}}(s)\,.
 \end{eqnarray}
The $Q^2$ dependence variables $\tau_2$ and $\tau_3$ in (\ref{eq:fi-nonsinglet}) are defined as
\begin{equation}
\tau_2 \equiv \frac{1}{(4\pi)^2}\int_{Q_0^2}^{Q^2}\alpha_s^2(Q'^2)d \ln Q'^2\,,
\end{equation}
and
\begin{equation}
\tau_3 \equiv \frac{1}{(4\pi)^3}\int_{Q_0^2}^{Q^2}\alpha_s^3(Q'^2)d \ln Q'^2\,.
\end{equation}
The leading-order nonsinglet splitting function is given by $\Phi_{\text{NS}}^{\text{LO}}(s)$ in the Laplace space  $s$,
\begin{eqnarray}
\Phi_f^{\text{LO} } = 4 - \frac{8}{3}\left(\frac{1}{s + 1} + \frac{1}{s + 2} + 2\left(\gamma_E + \psi (s + 1)\right)\right)\,, \nonumber  \\
\end{eqnarray}
where $\gamma_E=0.577216\cdots$ is the Euler's constant and $\psi(s) = d \ln \Gamma(s)/ds$ is the digamma function. The next-to-leading order and next-to-next-to-leading order splitting functions $\Phi_{qq}^{\text{NLO}}$ and $\Phi_{qq}^{\text{NNLO}}$ are too lengthy to be included here and we list them in Appendix {\bf A}.

In summary, there are various numerical and analytical methods to solve the DGLAP evolution equations to obtain quark and gluon
structure functions.
In Ref.~\cite{Schoeffel:1998tz}, a Laguerre polynomials method is presented to solve the DGLAP equations. The method is based on the expansion of the PDFs and splitting functions in term of Laguerre polynomials, which reduces DGLAP integro-differential equations to a set of ordinary differential equations.
A Taylor series expansion method to solve the evolution equations is also presented in Ref.~\cite{Devee:2012zz} up to NNLO for the small value of Bjorken-$x$.
Among various methods of solving the DGLAP equations,  M. Hirai et al~\cite{Hirai:1997gb} employed a brute-force
method~\cite{Miyama:1995bd} for the spin-independent case. They have also investigated the Laguerre method to solve the DGLAP evolution equation for this case~\cite{Kumano:1992vd,Kobayashi:1994hy}. The Mellin-transformation method for solving the Q$^2$ evolution equations of DGLAP evolution equations are studied in details in Refs.~\cite{Gluck:1989ze,Graudenz:1995sk,Blumlein:1997em,Blumlein:2000hw,Stratmann:2001pb}. Ref.~\cite{Vogt:2004ns} introduces a computer code entitled {\tt QCD-PEGASUS}, which allows one to perform DGLAP evolution up to NNLO in QCD using the Mellin transformation method. One can also obtain the numerical solution of the DGLAP evolution equations for the evolution of unpolarized and polarized parton distributions of hadrons in $x$ space using the QCD evolution program called {\tt QCDnum}~\cite{Botje:2010ay}.

As we mentioned, we have employed the Laplace transforms method for the QCD analysis of neutrino-nucleon structure functions. The accuracy with which one can be obtained from the analytical solution using the inverse Laplace transforms technique is found to be better than one part in 10$^{4-5}$~\cite{Block:2009en}.
The solutions which are obtained, resulting from the iterated expressions, have enough accuracy
to be competed with the results of other groups. This method can be extended to a higher
order approximation and is applicable even when we encounter heavy quarks contributions where transition to heavier active
flavors is allowed. In this manuscript we have shown that the methods of Laplace transforms technique are also the reliable and alternative schemes to solve these equations, analytically. One can conclude that the Laplace transform method is sound and can be used as an
alternative to various other methods for the solution of DGLAP equations.

The next section introduces the theoretical perspectives on how the Jacobi polynomial approach can be used to extract the nonsinglet structure functions $xF_3(x, Q^2)$ from the solution of nonsinglet DGLAP equations at NNLO.
It is important to note that this type of solution of the DGLAP equations is capable of evolving structure functions at any order of $x$ and Q$^2$.

%
%

\subsection{ The Jacobi polynomial approach }\label{Jacobi}

In the charged-current neutrino DIS processes, the neutrino $\nu (\bar{\nu})$ scatters off a quark inside the nucleon via the exchange of a virtual $W^\pm$ boson.
The nonsinglet structure function $xF_3(x, Q^2)$, associated with the parity-violating weak interaction, represents the momentum density of valence quarks.
At the LO approximation the $xF_3(x, Q^2)$ structure function, which is a combination of  valence quark densities, can be written as
\begin{eqnarray}
	xF_3^{\nu N} &=& x u_v(x) + x d_v(x)+ 2 \, x s(x) - 2 \, xc(x)\, ,  \nonumber  \\
	xF_3^{\bar{\nu} N} &=& x u_v(x) + x d_v(x)- 2 \, x s(x) + 2 \, x c(x) \;,
\end{eqnarray}
where the $d_v \equiv d - \bar d$ and $u_v  \equiv u - \bar u$ combinations are the proton
valence densities.
The asymmetry of the $ s(x) - c(x)$ distribution leads to the result  in which $xF_3^{\nu N} \neq  xF_3^{\bar{\nu} N}$.
The data reported by the CCFR~\cite{Seligman:1997mc}, NuTeV~\cite{Tzanov:2005kr} and CHORUS~\cite{Onengut:2005kv} collaborations present the average of the neutrino and
anti-neutrino distributions. Therefore, one can write the nucleon structure function as
\begin{equation}
	xF_3(x) = \frac{xF_3^{\nu N} + xF_3^{\bar{\nu} N}}{2} = xu_v(x) + xd_v(x) \,.
\end{equation}

For the present analysis, we use the following standard parametrizations at the input scale Q$_0^2$ = 4 GeV$^2$ for all valence distributions, $xu_v$ and $xd_v$:
\begin{equation}\label{eq:xuvQ0}
	x \, u_v = {\cal N}_u x^{\alpha_{u_v}}(1-x)^{\beta_{u_v}}( 1 + \gamma_{u_v} x^{0.5} + \eta_{u_v} x )\,,
\end{equation}
\begin{eqnarray} \label{eq:xdvQ0}
	x \, d_v  & = & \frac{{\cal N}_d} {{\cal N}_u}(1 - x)^{\beta_{d_v}} xu_v\,    \nonumber  \\
&=& {\cal N}_d x^{\alpha_{u_v}}(1-x)^{\beta_{u_v}+\beta_{d_v}}( 1 + \gamma_{u_v} x^{0.5} + \eta_{u_v} x )\,.	\nonumber \\
\end{eqnarray}
Since there are not enough data to  constrain the fit parameters sufficiently, especially for the medium values of Bjorken scaling $x$, then we reduce the number of free parameters by considering the above distribution for the $d$-valence distribution $xd_v$. Consequently, the free parameters \{$\alpha$, $\gamma$ and $\eta$ \} would be the same both for $x \, u_v$ and $x \, d_v$ distributions. As a result, $\beta_{u_v} + \beta_{d_v}$ will be set as a power of $(1 - x)$ for $x \, d_v$, because at high $x$ this PDF is becoming small and we attempt to constrain it at this region. In addition, this parametrization give $x \, d_v$ sufficient flexibility at medium and large value of $x$.
Normalization factors ${\cal N}_u$ and ${\cal N}_d$ are given by
\begin{eqnarray}\label{eq:noal}
	{\cal N}_u &=& 2/ \left(B[\alpha_{u_v}, \beta_{u_v} + 1] + \eta_{u_v}B[\alpha_{u_v} +1, \beta_{u_v}+1]\right.  \nonumber \\ & + & \left. \gamma_{u_v} B[\alpha_{u_v} + 0.5, \beta_{u_v} + 1]\right) \,,   \nonumber  \\ \\
	{\cal N}_d &=& 1/\left(B[\alpha_{u_v}, \beta_{d_v}+\beta_{u_v} + 1]\right.    \nonumber  \\
	& +&\left.   \eta_{u_v} B[\alpha_{u_v} + 1, \beta_{d_v} + \beta_{u_v} + 1]\right.    \nonumber  \\ & + & \left. \gamma_{u_v} B[\alpha_{u_v} + 0.5, \beta_{d_v} + \beta_{u_v} + 1]\right)\,,
\end{eqnarray}
where $B$ is the well-known Euler beta function. \\
Defining $x = e^{-\nu}$ and moving to the Laplace space $s$, via the following transformations
\begin{equation}
	u_v(s) = {\cal L}[e^{-\nu} u_v(e^{-\nu}); s] \,,
\end{equation}
\begin{equation}
	d_v(s) = {\cal L}[e^{-\nu} d_v(e^{-\nu}); s] \,,
\end{equation}
one can obtain the valence distributions presented in (\ref{eq:xuvQ0}) and (\ref{eq:xdvQ0}) in the Laplace space as

\begin{eqnarray}
	&& u_v(s)=2 (B[\alpha_{u_v} + s, \beta_{u_v} + 1] + \eta_{u_v} B[\alpha_{u_v} + s + 1, \beta_{u_v} + 1]   \nonumber  \\
	&& + \gamma_{u_v} B[\alpha_{u_v} + s + 0.5, \beta_{u_v} + 1])/(B[\alpha_{u_v}, \beta_{u_v} + 1]     \nonumber   \\
	&&+ \eta_{u_v} B[\alpha_{u_v} + 1, \beta_{u_v} + 1] + \gamma_{u_v} B[\alpha_{u_v}  +0.5, \beta_{u_v} + 1]),
\end{eqnarray}
and,
\begin{eqnarray}
	&& d_v(s)=(B[\alpha_{u_v} + s, \beta_{d_v}+\beta_{u_v} + 1]   \nonumber  \\
	&&+\eta_{u_v} B[\alpha_{u_v} + s + 1, \beta_{u_v} + \beta_{d_v} + 1]   \nonumber \\
	&&+ \gamma_{u_v} B[\alpha_{u_v} + s + 0.5, \beta_{u_v} + \beta_{d_v} + 1])/(B[\alpha_{u_v}, \beta_{d_v} + \beta_{u_v} + 1]   \nonumber \\
	&&+ \eta_{u_v} B[\alpha_{u_v} + 1, \beta_{u_v}+\beta_{d_v}+1]   \nonumber   \\
	&& +\gamma_{u_v} B[\alpha_{u_v} + 0.5, \beta_{u_v} + \beta_{d_v} + 1]).
\end{eqnarray}

Having the NNLO contributions of the Wilson coefficient functions in Laplace $s$-space, one can construct the $xF_3(x)$ structure function up to three-loops order.
Finally the nonsinglet structure function $xF_3(x, Q^2)$ in Laplace $s$-space, up to the NNLO approximation, can be written as
\begin{eqnarray}\label{eq:xf3-laplace-space}
	&& F_3(s, \tau) = e^{\tau \, \Phi_{\text{ NS}}(s)} (~ u_v(s) + d_v(s) ~)  \nonumber  \\
	&& \times\big( 1 + \tau/(4\pi) ~ C_3^{(1)}(s) + (\tau/(4\pi))^2 ~ C_3^{(2)}(s)  \big ) \,,  \nonumber  \\
\end{eqnarray}
where the coefficients $C_3^i$ are the common Wilson coefficients in Laplace space. One can easily determine these NNLO coefficient functions in Laplace $s$ space using the NNLO results derived in Refs.~\cite{Moch:1999eb,vanNeerven:1999ca}.
The corresponding NLO and NNLO coefficient functions in the Laplace  space can be found in Appendix {\bf A}.

As we mentioned, the present analysis is based on the Jacobi polynomial technique of reconstruction of the structure function from its Laplace moments.
The extracted results for the DGLAP evolution equations (\ref{eq:nonsinglet-laplace-space}) and the structure functions in Laplace space (\ref{eq:xf3-laplace-space}) are used as input to obtain the $x$ and Q$^2$ evolution of the $xF_3(x, Q^2)$ structure function. The Jacobi polynomials approach is also used to facilitate the analysis.
The method of Jacobi polynomials QCD analysis of proton structure functions are discussed in details in Ref.~\cite{Khorramian:2009xz} and successfully applied in the process of the fits of DIS data, so we explain only a brief outline here. In this method, each given structure function may be reconstructed in a form of the series as follows
\begin{eqnarray}\label{eq:xF_3Jacobi}
x F_3(x, Q^2)  & = & x^{\beta}(1 - x)^{\alpha} \, \sum_{n=0}^{N_{max}} \, \Theta_n^{\alpha, \beta}(x) ~ a_n(Q^2)\, \nonumber \\
& = & x^{\beta}(1 - x)^{\alpha} \, \sum_{n=0}^{N_{max}} \, \Theta_n^{\alpha, \beta}(x) \nonumber \\
&\times & \sum_{j=0}^n \, c_j^{(n)}{(\alpha, \beta)} \, {{\cal L}} [xF_3, s=j + 1] \,,
\end{eqnarray}
where  $N_{\rm max}$ is the number of polynomials which normally sets to 7 or 9 and, $a_n(Q^2)$ are the Jacobi moments.
Form Eq.~\ref{eq:xF_3Jacobi}, one can conclude that the use of Jacobi polynomials has this
advantage to allow us to factor out the essential part of the $x$-dependence
of the structure function into a weight function $x^{\beta}(1 - x)^{\alpha}$ and the Q$^2$ dependence is contained in the Jacobi moments. On the right-hand side of the above equation, the ${{\cal L}} [xF_3, s=j + 1]$ are the Laplace transformation of the structure functions.

The $\Theta_n^{\alpha, \beta}(x)$ in Eq.~(\ref{eq:xF_3Jacobi}) are the Jacobi polynomials with the following expansion,
\begin{equation}\label{eq:Jacobi-polynomials}
\Theta_n^{\alpha, \beta}(x) = \sum_{j = 0}^{n} \, c_j^{(n)}(\alpha, \beta) \, x^j \,,
\end{equation}
where $c_j^{(n)}(\alpha, \beta)$ are the coefficients which are expressed through $\gamma$-functions. The $\alpha$ and $\beta$ parameters are fixed to 3 and 0.5, respectively.

In the results of our previous analysis~\cite{Khorramian:2009xz}, in which we obtained with the fixed weight function of the Jacobi polynomials reconstruction formula, namely $x^{0.5} (1 - x)^3$, we found that by choosing the set of \{N$_{max}$=9,  $\alpha$=3, $\beta$=0.5\} one can achieve the optimal convergence of these series throughout the kinematic region constrained by the data. In the present analysis, we found that the selected values to be sufficient to achieve the fastest convergence of the series on the right-hand side of Eq.~(\ref{eq:xF_3Jacobi}) and to reconstruct the $xF_3$ structure function with the required accuracy.
We have checked that the results of our NLO and NNLO fits are almost non-sensitive to the changes of, N$_{max}$ = 10 to N$_{max}$ = 6, which was considered in the process of the NNLO fit. Consequently, we fixed this parameter to N$_{max}$=9 as our previous analysis for the $F_2^{\rm NS} (x, Q^2)$ structure function~\cite{Khorramian:2009xz}.
One can conclude that the selected form of the weight function $x^{0.5} (1 - x)^3$ is similar to the $x$-shape of the non-singlet structure function itself~\cite{Kataev:1999bp,Brodsky:1973kr}.
However, one can consider the $\alpha$ and $\beta$ as free parameters in the fit.
Considering the problem of minimization of the dependence of the fits results to the $\alpha$ and $\beta$, we found several values for these parameters.
Overall we found that considering the obtained results and in view of the stability of the results of NLO and NNLO analyses to the selected choice, $\alpha$ = 3 and $\beta$ = 0.5, we considered this minimum as the physical one.

The Jacobi polynomials satisfy the following orthogonality relation with the weight function $x^{\beta} (1 - x)^{\alpha}$,
\begin{equation}
\int_0^1 dx \, x^{\beta} (1 - x)^{\alpha} \, \Theta_{k}^{\alpha, \beta}(x) \, \Theta_{l}^{\alpha, \beta}(x) = \delta_{k,l} \,.
\end{equation}\label{eq:orthogonality-relation}
The extracted $x F_3(x, Q^2)$ structure function can be used for the QCD analysis of the nonsinglet structure function $xF_3(x, Q^2)$ measured in CCFR~\cite{Seligman:1997mc} and NuTeV~\cite{Tzanov:2005kr} experiments at Fermilab and recent neutrino oscillation search by the CHORUS collaboration~\cite{Onengut:2005kv} at CERN.
These data can provide a precise experimental source to determine the valence-quark distributions, $x u_v (x, Q^2)$ and $x d_v (x, Q^2)$.
It is also worth mentioning that one can include the heavy-flavor contributions to the above nonsinglet charged-current $\nu$-nucleon DIS structure functions~\cite{Behring:2015roa,Blumlein:2016xcy,Laenen:1992zk,Riemersma:1994hv}.

%
\section{ Global analyses of valence-quarks densities }\label{global-PPDFs}
%

%
%
\subsection{ Choice of data sets }
The recent measurements of the CCFR, NuTeV and CHORUS collaborations provide the most precise up to now
experimental results for the structure functions  $xF3(x, Q^2)$ of the DIS of neutrinos and antineutrinos on nucleons.
The data for the charged-current structure functions $xF_3(x, Q^2)$ used in our analysis are listed in Table.~\ref{table:xf3data}. The $x$ and $Q^2$ ranges, the number of data points and the related references are also listed in this table.
\begin{table*}[htb]
	\begin{tabular}{|c|c|c|c|c|c|c|}
		\hline
		{\tt Experiment} & $x$    & Q$^2$ & Number of data points  & Reference  &  $\chi^2$-{\tt NLO}       &  $\chi^2$-{\tt NNLO}    \\      \hline   \hline
		{\tt CCFR}	 & $0.0075 \leq x \leq 0.75$    & $1.3 \leq Q^2 \leq 125.9 $ & 116 & \cite{Seligman:1997mc}& $285.37$  & $270.26$  \\
		{\tt NuTeV}	 &   $0.015 \leq x \leq 0.75$  &$3.162 \leq Q^2 \leq 50.118$  & 64 & \cite{Tzanov:2005kr} & $209.09$ & $207.84$   \\
		{\tt CHORUS}	 &  $0.02 \leq x \leq 0.65$   &$2.052 \leq Q^2 \leq 81.55$  & 50 & \cite{Onengut:2005kv} & $117.25$  &  $111.51$  \\   \hline
	\end{tabular}
	\caption{ Published data points for charged-current structure functions $xF_3(x, Q^2)$ used in the present global fit. The $x$ and $Q^2$ ranges, the number of data points and the related references are also listed. The $\chi^2$ values corresponding to each of the three data sets for each of the NLO and NNLO fits are also presented. \label{table:xf3data} }
\end{table*}
The CCFR~\cite{Seligman:1997mc} and NuTeV~\cite{Tzanov:2005kr} collaborations at the Fermilab use an iron target in their neutrino deep inelastic scattering experiments, corrected to an isoscalar target, and cover much the same kinematic range of momentum fraction $x$ but the CCFR covers slightly higher Q$^2$. At high values of $x$, the predictions are mainly determined by the valence up quark distribution, which is very well constrained by the neutral-current DIS structure
function data. We also include the recent data form CHORUS~\cite{Onengut:2005kv} collaboration
which are  taken from a lead target and cover a similar range in $x$ in comparison with CCFR.
The NuTeV data seems to be more precise. In practice, we find the high-$x$ NuTeV data very difficult to fit so that lead to higher values of $\chi^2$.
Different theoretical treatment of nuclear effects could make a difference at small and large values of momentum fraction $x$.
NuTeV indicates that neutrino scattering favors smaller nuclear effects at high-$x$ than are found in charged-lepton
scattering experiments~\cite{Tzanov:2005kr}. New theoretical calculations in the shadowing region, in which the nuclear correction has Q$^2$
dependence, imply that one can ignore heavy target data for $x > 0.1$ in the fit.
Since, we mostly focused on the method of Laplace transform and Jacobi polynomials in term of speed and accuracy, we preferred in using the neutrino-nucleon data over the whole $x$ range, rather than just for $x > 0.1$.

\subsection{The method of minimization}

To determine the best values of the fit at next-to-leading and next-to-next-to-leading orders, we need to minimize the $\chi^2$ with respect to five free input valence-quark distribution parameters of Eqs.~(\ref{eq:xuvQ0}, \ref{eq:xdvQ0}) including the QCD cutoff parameter $\Lambda_{\rm{\overline {MS}}}^{(4)}$. In our analysis, the global goodness-of-fit procedure follows the usual method with $\chi^2 (p)$ defined as
\begin{equation} \label{chi2}
\chi^2 (p) = \sum_{\text{i = 1}}^{n^{\text{data} }} \frac{ (~ D_i^{\text{data} } - T_i^{\text{theory} } (p) ~ )^2}{ ( \sigma_i^{\text{data} } )^2 } \,,
\end{equation}
where $p$ denotes the set of six independent free parameters in the fit and $n^{\rm data}$ is the number of data points included, so $n^{data}$ = 230 in our work.
The widely-used CERN program library {\tt MINUIT}~\cite{CERN-Minuit} is applied to obtain the best parametrization of the valence-quark PDFs.
The experimental errors are calculated from systematic and statistical errors added in quadrature, $\sigma_i^{\text{data}} = \sqrt{(\sigma_i^{\text{sys }})^2 + (\sigma_i^{\text{stat}})^2 }$.
The $\chi^2$ values corresponding to each individual data set for each of the NLO and NNLO fits are presented in Table.~\ref{table:xf3data}.
The largest contributions to $\chi^2$ arise from the NuTeV deep inelastic neutrino-nucleon structure functions,
with smaller contributions from low-$x$ CHORUS data, and medium-$x$ CCFR data.
From the Table.~\ref{table:xf3data}, one can conclude that the precise NuTeV data set is very difficult to fit so that led to higher values of $\chi^2$.
For this data set we obtained $\chi^2/n^{\rm data} = 209.09/64$ for the NLO analysis and $\chi^2/n^{\rm data} = 207.84/64$ for the NNLO one.
The motivation for using the NuTeV data set in our analysis comes mainly from adding a new and up-to-date data set for the neutrino-nucleon scattering structure function.

%
\subsection{  Uncertainties on input PDFs  }
%
Now we are in a situation to present our method for the calculation of the valence-quark PDFs uncertainties and error propagation from experimental data points. To obtain the uncertainties in global PDF analyses, there are well-defined procedures for propagating experimental uncertainties on the fitted data points through to the PDF uncertainties. Here, we use the Hessian method (or error matrix approach) \cite{Pumplin:2001ct}, which is based on linear error propagation and involves the production of eigenvector PDF sets suitable for convenient use by the end user.
Originally, the Hessian method was  used in MRST~\cite{Martin:2003sk} and MSTW08~\cite{Martin:2009iq} analyses and we also applied this approach in our previous works~\cite{AtashbarTehrani:2012xh,Khanpour:2016pph,Shahri:2016uzl}. Therefore, hereinafter we concentrate on this method.
Following that, an error analysis can be done by using the Hessian matrix, which is obtained by running the
CERN program library MINUIT~\cite{CERN-Minuit}.
The most commonly applied Hessian approach, which is based on the covariance matrix diagonalization, provides us a simple and
efficient method for calculating the uncertainties of PDFs.
The basic assumption of the Hessian approach is a quadratic expansion
of the global goodness-of-fit quantity, $\chi^2_{\text{ global}}$, in the fit parameters $a_i$ near the global minimum,
\begin{equation} \label{chi2-2}
\Delta \chi^2_{\text{ global}} \equiv  \chi^2_{\text{global}}  - \chi^2_{\text min} = \sum_{\text{ i, j=1}}^n (a_i - a_i^0) \, H_{\rm ij} \,   (a_j - a_j^0) \,,
\end{equation}
where $H_{\rm ij}$ are the elements of the Hessian matrix and $n$ stands for the  number of parameters in the global fit.

The uncertainty on a partonic distribution function $f(x, a_i)$ is then given by
\begin{eqnarray}\label{deltak2}
\delta f(x, a_i)& = &    \nonumber \\
\biggl[\Delta \chi^2 & & \sum_{i, j}^n  \left ( \frac{\partial f(x, a)}{\partial a_i}
\right) _{a = \hat{a}}  H_{ij}^{-1}  \left ( \frac{\partial f(x, a)}{\partial a_j} \right)_{a = \hat{a}} \biggr]^{1/2}, \nonumber \\
\end{eqnarray}
where $a_i$ stand for the fit parameters in the input valence distributions (\ref{eq:xuvQ0}, \ref{eq:xdvQ0}), and $\hat a$ indicates the number of parameters which make an extreme value for the related derivative.
Running the CERN program library {\tt MINUIT}, the Hessian or covariance matrix elements for six free parameters in  our NLO and NNLO global analysis are given in Tables~\ref{covmat-matrixNLO} and \ref{covmat-matrixNNLO}, respectively.
The uncertainties of PDFs are estimated using these Hessian matrix  explained  and their values at higher Q$^2$ (Q$^2$>Q$_0^2$) are calculated using the DGLAP evolution equations.
\begin{table*}[htb]
	\renewcommand{\arraystretch}{1.30}
	\centering
	{\footnotesize
		\begin{tabular}{||c||c|c|c|c|c|c||}
			\hline \hline
			& $\alpha_{u_v}$ & $\beta_{u_v}$ &$\gamma_{u_v}$ & $\eta_{u_v}$ & $\beta_{d_v}$ & $\Lambda_{\overline{\rm MS}}^{(4)}$       \\
			
			\hline \hline
			$\alpha_{u_v}$       & 2.193 $\times 10^{-6}$ &  &  &  & &  \\
			\hline
			$\beta_{u_v}$       & -2.880 $\times 10^{-5}$  & 5.834 $\times 10^{-4}$&  &  &  & \\
			\hline
			$\gamma_{u_v}$       &-1.563 $\times 10^{-3}$ & 2.931 $\times 10^{-2}$ & 1.928 &  & &  \\
			\hline
			$\eta_{u_v}$       &1.576 $\times 10^{-3}$  &-3.039$\times 10^{-2}$ &   -1.749&   2.085              &  & \\
			\hline
			$\beta_{d_v}$       &-3.151 $\times 10^{-4}$ &  5.736 $\times 10^{-3}$  & 0.323  & -0.334  &6.162 $\times 10^{-2}$   &   \\
			\hline
			$\Lambda_{\overline{\rm MS}}^{(4)}$& -2.841 $\times 10^{-5}$ &5.311 $\times 10^{-4}$ & 3.032 $\times 10^{-2}$ &  -3.083 $\times 10^{-2}$   & 5.872 $\times 10^{-3}$  & 6.267 $\times 10^{-4}$ \\
			\hline\hline
		\end{tabular}
	}
	\caption[]{ The covariance matrix for the 5 + 1 free parameters in the NLO fit. }
	\label{covmat-matrixNLO}
\end{table*}
\begin{table*}[htb]
	\renewcommand{\arraystretch}{1.30}
	\centering
	{\footnotesize
		\begin{tabular}{||c||c|c|c|c|c|c||}
			\hline \hline
			& $\alpha_{u_v}$ & $\beta_{u_v}$ &$\gamma_{u_v}$ & $\eta_{u_v}$ & $\beta_{d_v}$ & $\Lambda_{\overline{\rm MS}}^{(4)}$       \\
			
			\hline \hline
			$\alpha_{u_v}$       & 7.065 $\times 10^{-6}$ &  &  &  & &  \\
			\hline
			$\beta_{u_v}$       & -8.544 $\times 10^{-5}$  & 1.154 $\times 10^{-3}$&  &  &  & \\
			\hline
			$\gamma_{u_v}$       &-5.302 $\times 10^{-3}$ & 6.776 $\times 10^{-2}$ & 4.448 &  & &  \\
			\hline
			$\eta_{u_v}$       &3.151 $\times 10^{-3}$  &-4.148$\times 10^{-2}$ &   -2.562&   1.825              &  & \\
			\hline
			$\beta_{d_v}$       &-5.882 $\times 10^{-4}$ &  7.590 $\times 10^{-3}$  &  0.467  & -0.285  &5.164 $\times 10^{-2}$   &   \\
			\hline
			$\Lambda_{\overline{\rm MS}}^{(4)}$& -9.933 $\times 10^{-6}$ &1.278 $\times 10^{-4}$ & 7.899 $\times 10^{-3}$ &  -4.824 $\times 10^{-3}$   & 8.841 $\times 10^{-4}$  & 8.841 $\times 10^{-4}$ \\
			\hline\hline
		\end{tabular}
	}
	\caption[]{ As in Table~\ref{covmat-matrixNLO}, but for the NNLO fit. }
	\label{covmat-matrixNNLO}
\end{table*}

\begin{figure}
	\begin{center}
		\vspace{1cm}
		\resizebox{0.40\textwidth}{!}{\includegraphics{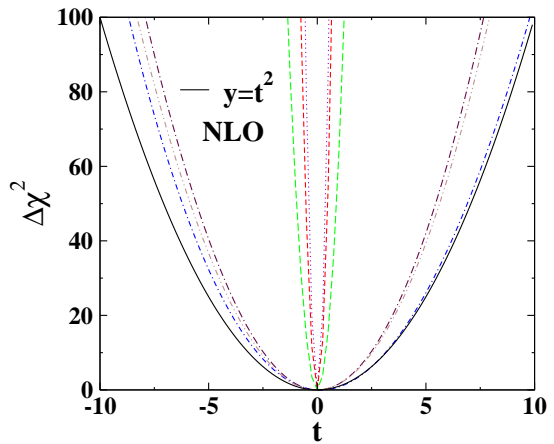}}  
		\caption{(Color online) $\Delta \chi^2$ as a function of $t$ defined in  Refs.~\cite{Khanpour:2016pph,AtashbarTehrani:2012xh,Martin:2002aw,Pumplin:2001ct} in the NLO approximation. The results correspond to some random sample of eigenvectors. The solid line correspond to the ideal case, $\Delta \chi^2_{\text{ global}} =  t^2$. }\label{fig:deltachinlo}
	\end{center}
\end{figure}
\begin{figure}
	\begin{center}
		\vspace{1cm}
		\resizebox{0.40\textwidth}{!}{\includegraphics{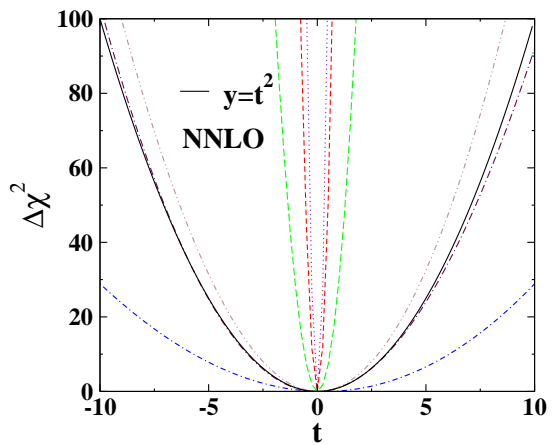}}  
		\caption{(Color online) As in Fig.~\ref{fig:deltachinlo}, but for the NNLO approximation. For comparison, we also display the Hessian approximation given by the quadratic form  $\Delta \chi^2_{\text{ global}} =  t^2$. }\label{fig:deltachinnlo}
	\end{center}
\end{figure}

%
\subsection{  Error propagation from experimental data  }
%
For the error calculations, we again follow the method presented in Refs.\cite{Khanpour:2016pph,Martin:2002aw,Pumplin:2001ct,Martin:2009iq,Pumplin:2009bb,Paukkunen:2014zia,Carrazza:2016htc}. In this method, one can work with the eigenvectors and eigenvalues of the covariance (or Hessian) matrix.
By having a set of appropriate fit parameters considered for the  valence-quark  PDFs which minimize the global $\chi^2_{\text{ global}}$ function, and introducing parton sets $s_k^\pm$, the parameter variation around the global minimum can be
expanded in a basis of eigenvectors and eigenvalues as
\begin{equation}
a_i (s_k^\pm)  = a_i^0  \pm  t \sqrt{\lambda_k} v_{\text {ik}}   \,,
\end{equation}
where $\lambda_k$ is the k$^{\text {th}}$ eigenvalue and $v_{\text {ik}}$ is the i$^{\text {th}}$ component of the orthonormal eigenvectors of the Hessian matrix. The parameter $t$ is adjusted to give the desired $T^2 = \Delta \chi^2_{\text{ global}}$ in which for the quadratic approximation we can set $t = T$, where $T$ is a tolerance parameter for the required confidence interval. The Hessian formalism used in this analysis provides a reliable and efficient method for error calculations.
In order to quantify the uncertainties of the physical predictions that depend on the PDFs,
one must choose the tolerance parameter $T$ to correspond to the region of acceptable fits.
It is worth mentioning that, various groups have different approaches for obtaining confidence level (\text{C.L.}) criteria for the value of $\chi^2$ in the goodness-of-fit test~\cite{Alekhin:2013nda,Accardi:2016qay,Dulat:2015mca,Abramowicz:2015mha,Harland-Lang:2014zoa,Jimenez-Delgado:2014twa} which comes from the quality of the experimental data sets they used in their fits.
In the results presented in our recent spin-dependant PDFs analysis~\cite{Shahri:2016uzl} as well as in our nuclear PDFs analysis~\cite{Khanpour:2016uxh}, we followed the standard parameter-fitting criterion and considered a 68\% ($1\sigma$) confidence level (\text{C.L.}) limit by the choice of tolerance $T = (\Delta \chi^2)^{1/2}$ = 1. In this paper, we again follow the standard parameter-fitting criterion considering $T = (\Delta \chi^2)^{1/2}$ = 1 for the 68\% (1$\sigma$) \text{C.L.}.
However, the actual value of $\Delta \chi^2$ depends on the number of parameters to be simultaneously determined in the fit~\cite{Martin:2009iq}.

To test the quadratic approximation of Eq.~(\ref{chi2-2}), we study the dependence of
$\Delta \chi^2_{\text{ global}}$ along some selected samples of eigenvector directions. The corresponding plots for the NLO and NNLO analysis are illustrated in Figs.~\ref{fig:deltachinlo} and \ref{fig:deltachinnlo}, respectively. The solid line is the quadratic
approximation given by Hessian method, $\Delta \chi^2_{\text{ global}} =  t^2$.
For some selected eigenvalues of Figs.~\ref{fig:deltachinlo} and \ref{fig:deltachinnlo}, the quadratic approximation works extremely well; however, for a few of eigenvalues it can deviate from the $\chi^2$ function. Nevertheless, in all the cases, we are able to
obtain a good description of the actual $\chi^2$ function. One can conclude that the error
of PDFs obtained using the well-known Hessian formalism will closely reflect the actual $\chi^2$
function determined by the experimental data.

The results of the present QCD analysis will be discussed in much more details in the next section.

\begin{figure}[htb]
	\begin{center}
		\vspace{1cm}
		\resizebox{0.450\textwidth}{!}{\includegraphics{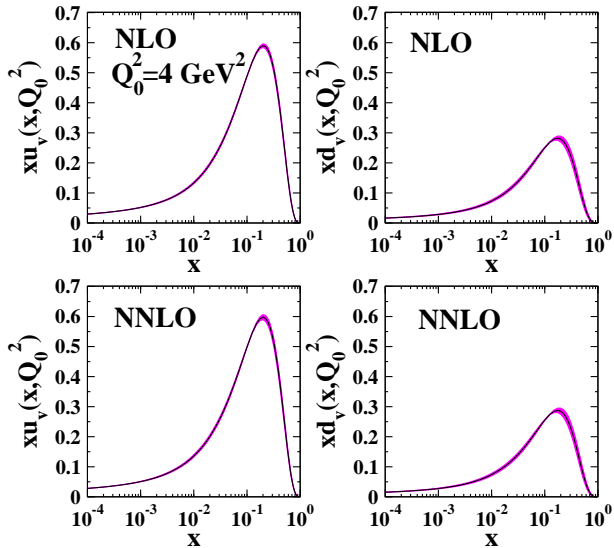}}  
		\caption{(Color online) The valence-quark parton densities $xu_v$ and $xd_v$ at the reference value Q$_0^2$ = 4 GeV$^2$ obtained from the NLO and NNLO global analyses. The corresponding $\Delta \chi^2$ = 1 uncertainty bands computed with the standard Hessian error matrix approach are also shown. }  \label{fig:partonQ0}
	\end{center}
\end{figure}

\begin{figure*}	
	\begin{tabular}{ll}
		\resizebox{0.55\textwidth}{!}{\includegraphics{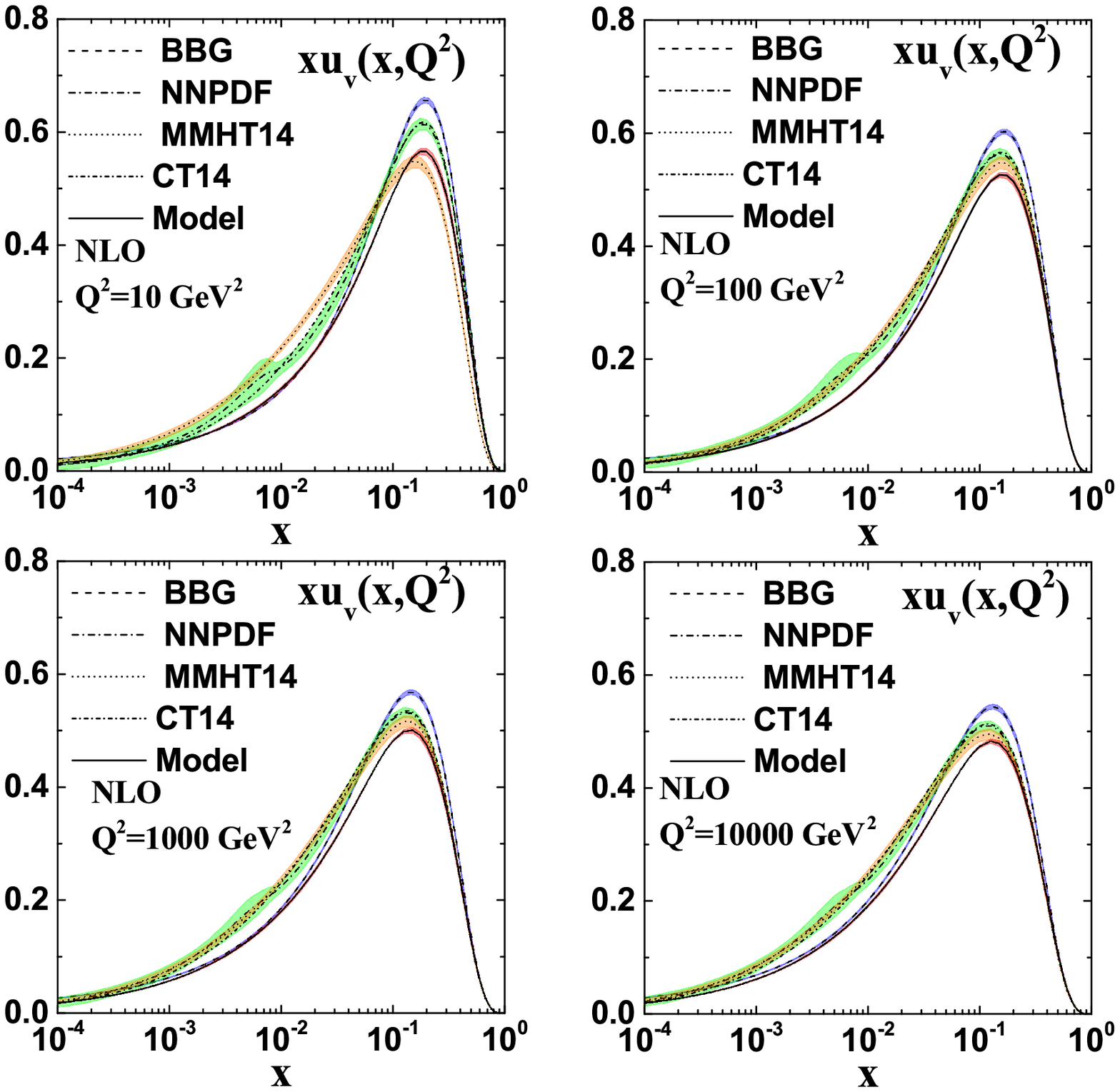}}   
		\resizebox{0.55\textwidth}{!}{\includegraphics{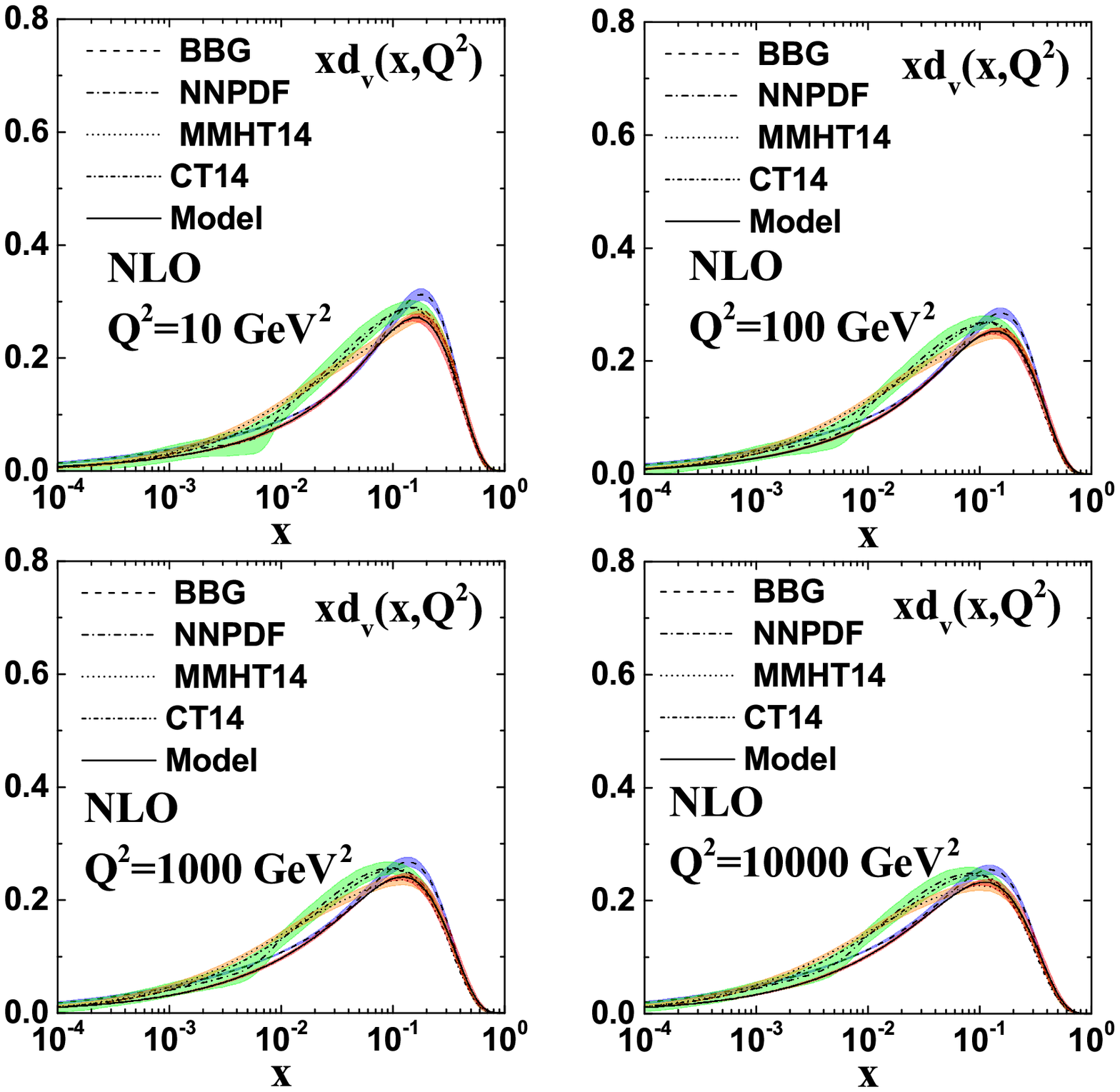}}   
	\end{tabular}
	\vspace{-40mm}
	\begin{center}
		\caption{(Color online) The up and down valence parton distributions $x u_v$ and $x d_v$ for some selected values of Q$^2$ = 10, 100, 1000,$ and $10000 GeV$^2$ in NLO order
			including their error bands. The dashed line is the {\tt BBG} model~\cite{Blumlein:2006be} , dashed-dotted is the {\tt NNPDF}~\cite{Ball:2012cx} model, short-dashed is the {\tt MMHT14}~\cite{Harland-Lang:2014zoa} model and short-dashed-dotted is the {\tt CT14}~\cite{Dulat:2015mca} model.}\label{fig:xuvandxdvnlo}
	\end{center}
\end{figure*}

\begin{figure*}
	\begin{tabular}{ll}
		\resizebox{0.55\textwidth}{!}{\includegraphics{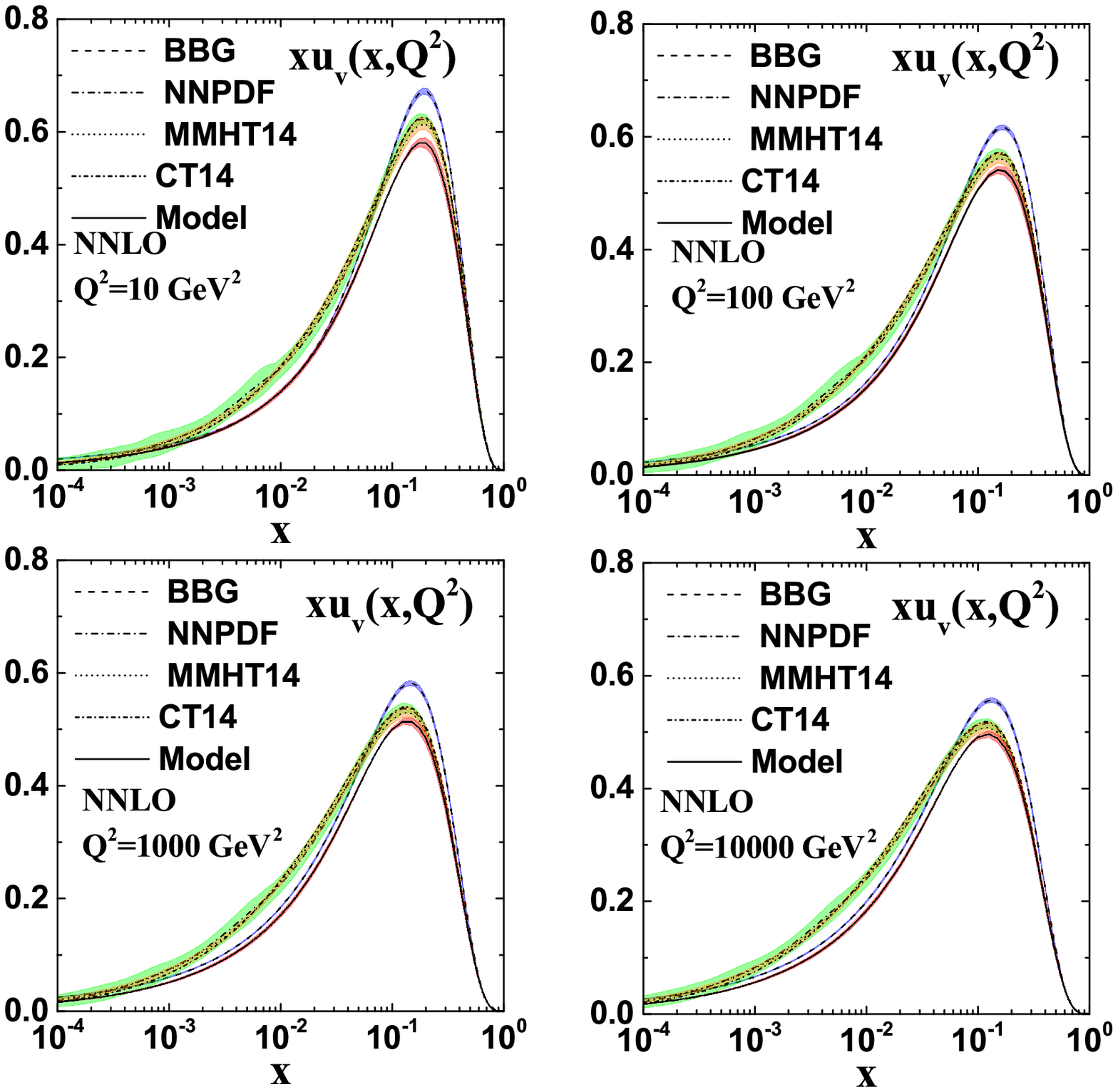}}   
		\resizebox{0.55\textwidth}{!}{\includegraphics{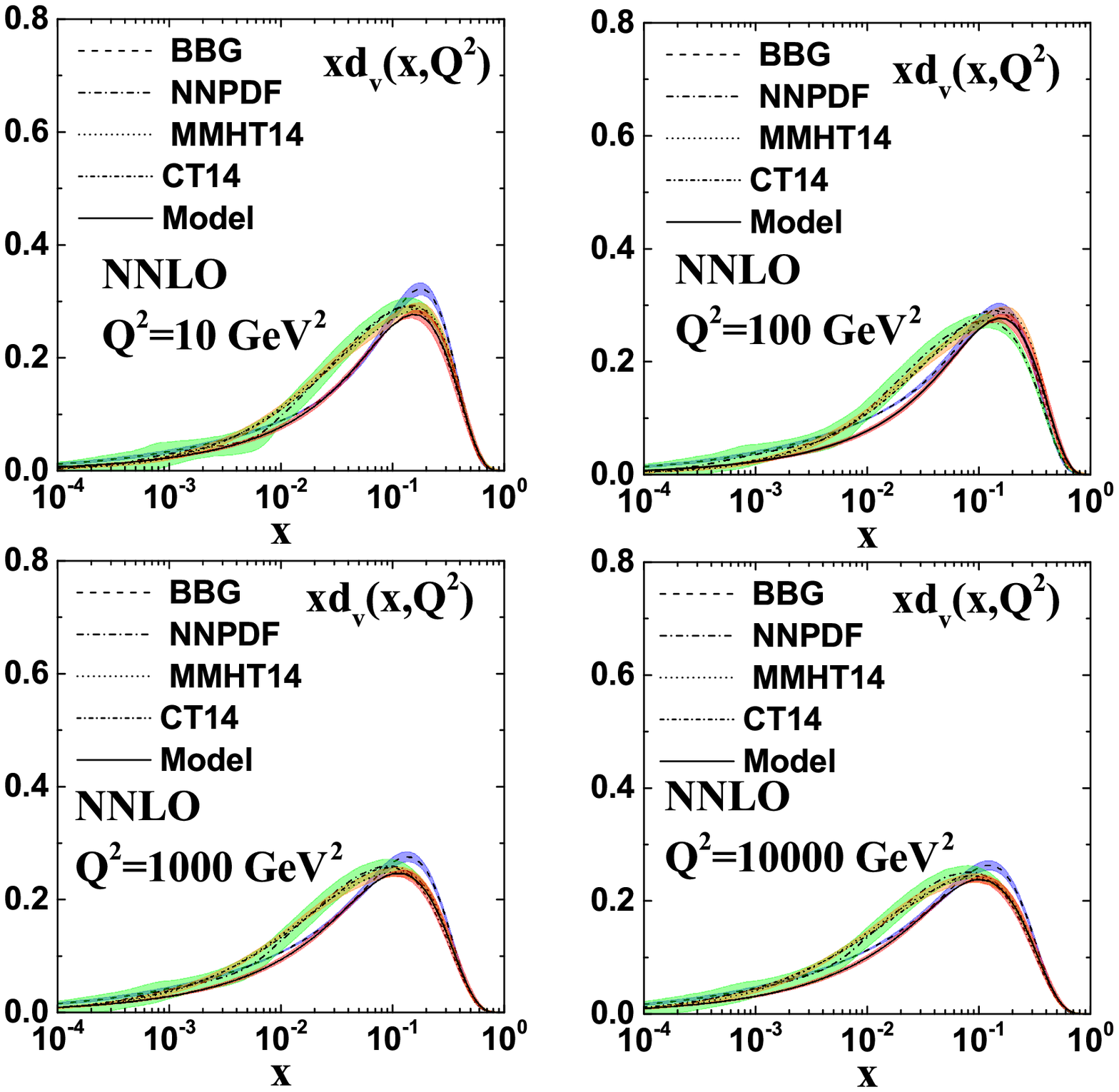}}   
	\end{tabular}
	\vspace{-40mm}
	\begin{center}
		\caption{(Color online) The up and down valence parton distributions $x u_v$ and $x d_v$ for some selected values of Q$^2$ = 10, 100, 1000,$ and $10000 GeV$^2$ in NNLO
			order including their error bands. The dashed line is the {\tt BBG} model~\cite{Blumlein:2006be}, dashed-dotted is the {\tt NNPDF}~\cite{Ball:2012cx} model, short-dashed is the {\tt MMHT14}~\cite{Harland-Lang:2014zoa} model and short-dashed-dotted is the {\tt CT14}~\cite{Dulat:2015mca} model.}\label{fig:xuvandxdvnnlo}
	\end{center}
\end{figure*}

\begin{figure}[htb]
	\begin{center}
		\vspace{1cm}
		\resizebox{0.50\textwidth}{!}{\includegraphics{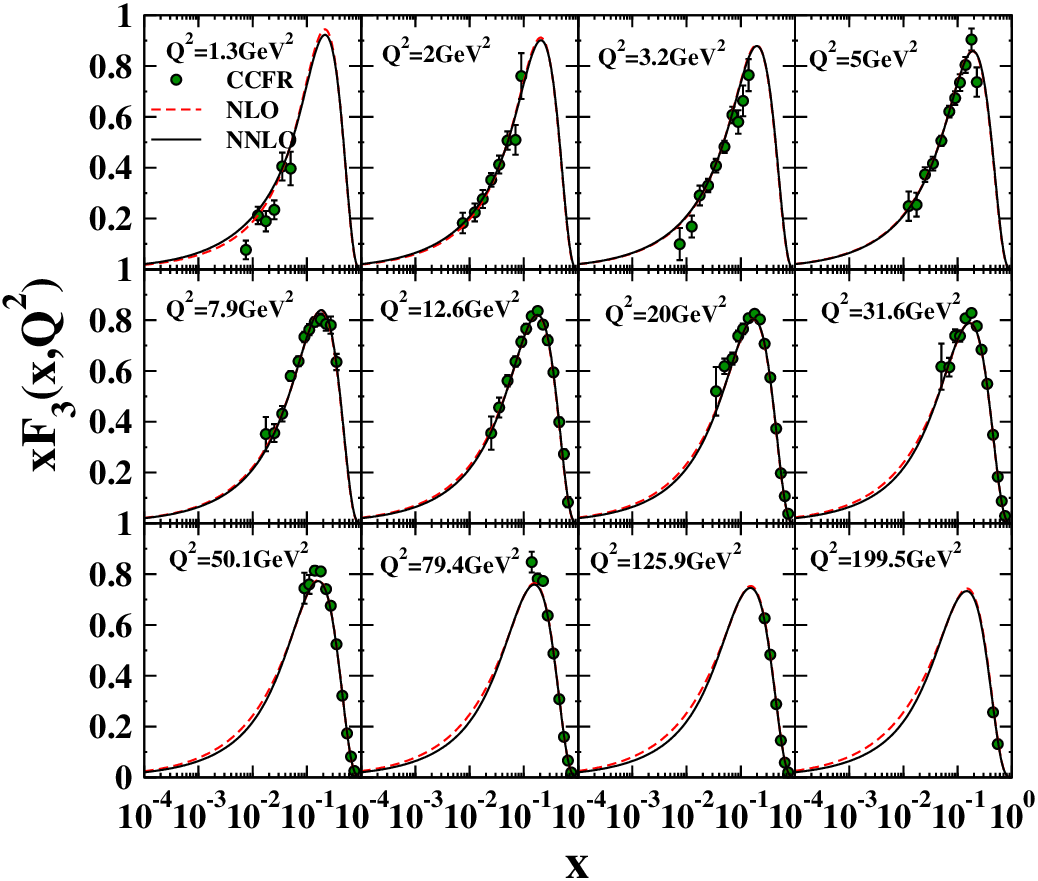}}  
		\caption{(Color online) The quality of the NLO and NNLO fits to the CCFR  antineutrino-initiated dimuon production. The dots represent the CCFR neutrino structure function measurements~\cite{Seligman:1997mc}. The solid curve represents our theoretical predictions as a function of $x$ and for some different values of Q$^2$.    } \label{fig:xf3ccfr}
	\end{center}
\end{figure}
\begin{figure}[htb]
	\begin{center}
		\vspace{1cm}
		\resizebox{0.50\textwidth}{!}{\includegraphics{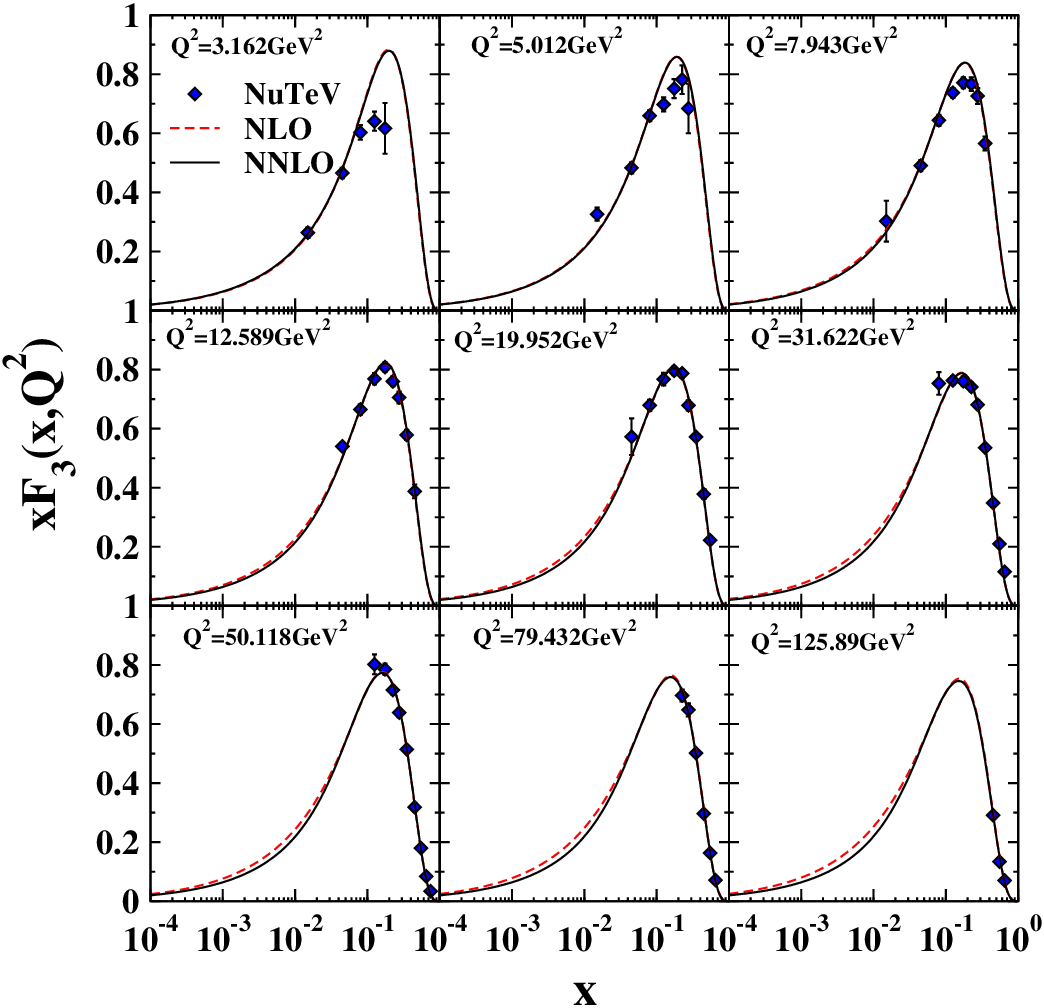}}  
		\caption{(Color online) The quality of the NLO and NNLO fits to the NUTEV antineutrino-initiated dimuon production. The dots represent the NuTeV neutrino structure function measurements~\cite{Tzanov:2005kr}. The solid curve represents our theoretical predictions as a function of $x$ and for some different values of Q$^2$.} \label{fig:xf3NuTev2006}
	\end{center}
\end{figure}
\begin{figure}[htb]
	\begin{center}
		\vspace{1cm}
		\resizebox{0.50\textwidth}{!}{\includegraphics{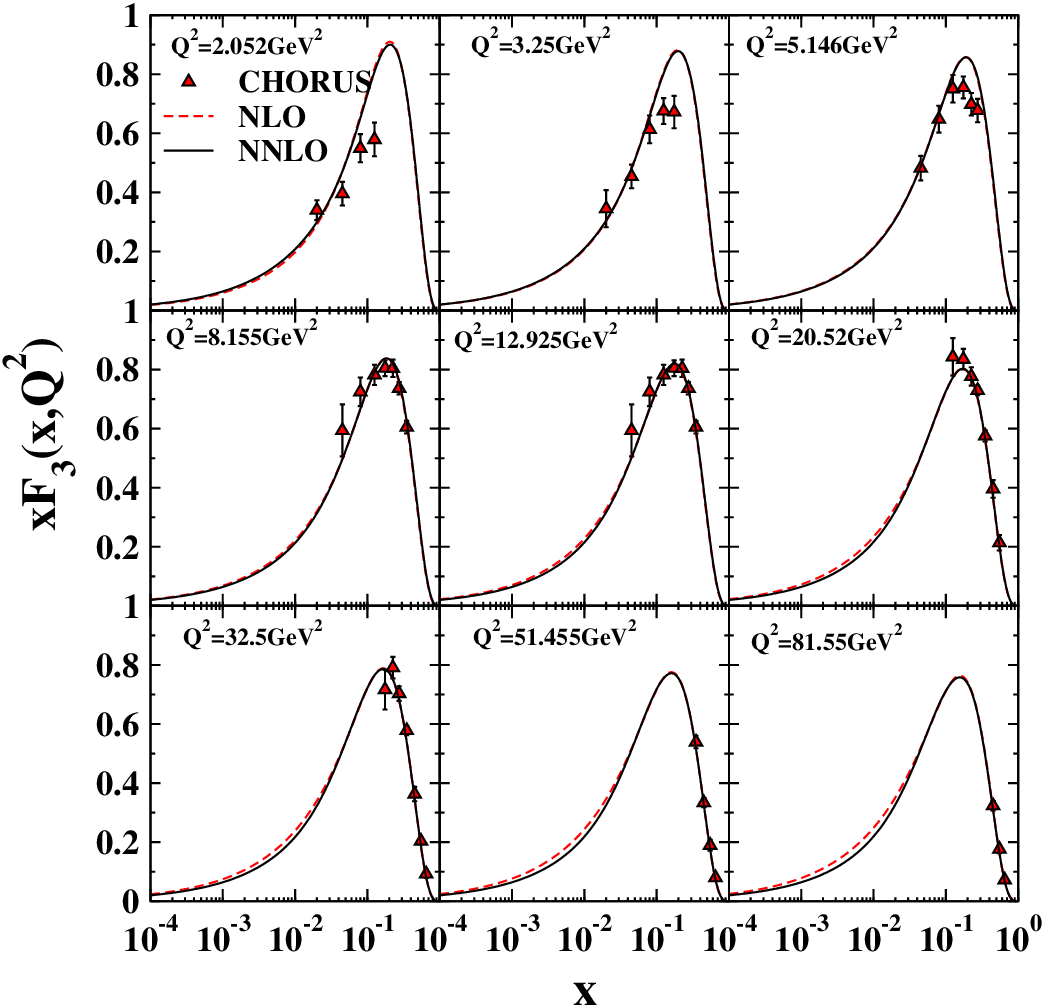}}  
		\caption{(Color online) The quality of the NLO and NNLO fits to the CHORUS antineutrino-initiated dimuon production. The dots represent the CHORUS neutrino structure function measurements~\cite{Onengut:2005kv}. The solid curve represents our theoretical predictions as a function of $x$ and for some different values of Q$^2$. }\label{fig:xf3chorus2006}
	\end{center}
\end{figure}

%
%
\section{Results and discussion}\label{Sec:Results}
In this section we present the results that have been obtained for the valence-quark densities, using the Laplace transformation technique and Jacobi polynomial approach, to solve analytically the DGLAP evolution equations and structure function.
For the input PDFs, we parameterized them with a standard form of distributions (\ref{eq:xuvQ0}, \ref{eq:xdvQ0}). Here, we determine the QCD scale $\Lambda_{\overline{\rm MS}}^{(4)}$ along with the parameters of the parton densities at
the initial scale $Q_0^2$. For each of the NLO and NNLO fits, the optimal values of the partonic distribution parameters shown in (\ref{eq:xuvQ0}, \ref{eq:xdvQ0}), along with the optimal values
of the QCD coupling $\alpha_s(Q_0^2)$ at the reference scale Q$_0^2$ = 4 GeV$^2$ are given in Table.~\ref{table:fitresults}. We believe that
we obtained good results for the PDFs parameters using the Laplace and Jacobi polynomials method.

Another important observation comes from the comparison of the values for $\alpha_s(M_Z^2)$ with the outcomes of the previous fits of the available DIS data~\cite{Martin:2009iq,Botje:2010ay,Breitweg:2001rq,Botje:1996hy,Alekhin:2011sk,Kniehl:2006bg,Agashe:2014kda}. We extracted the value of $\alpha_s(M_Z^2)=0.1161 \pm 0.00149$ for the strong coupling constant at the Z boson mass scale for the NLO approximation and $\alpha_s(M_Z^2)=0.1184 \pm 0.00047$ for the NNLO approximation. The result for the  $\alpha_s(M_Z^2)$ at NNLO are slightly higher than the results obtained from fitting to high-statistics charged-lepton structure function data alone.  In Fig.~\ref{fig:partonQ0}, we plot the valence-quark parton densities $xu_v$ and $xd_v$ at the input scale Q$_0^2$ = 4 GeV$^2$ obtain from the NLO and NNLO global analyses. The corresponding uncertainty bands are shown as well.

The up and down valence parton distributions, $x u_v$ and $x d_v$, at some selected values of Q$^2$ = 10, 100, 1000, and 10000 GeV$^2$ are plotted in Figs.~\ref{fig:xuvandxdvnlo} and \ref{fig:xuvandxdvnnlo} for NLO and NNLO analyses, respectively. The dashed-dashed-dotted and solid lines represents our model at NLO and NNLO approximations, respectively. The dashed line is the {\tt BBG} model~\cite{Blumlein:2006be}, dashed-dotted is the {\tt NNPDF}~\cite{Ball:2012cx} model, short-dashed represent the {\tt MMHT14}~\cite{Harland-Lang:2014zoa} model and short-dashed-dotted is the {\tt CT14}~\cite{Dulat:2015mca} model.
As can been seen from the plots, both $xu_v$ and $xd_v$ valence distributions from the {\tt BBG} model are slightly larger than other groups for $x \approx 0.1$.
The results from {\tt NNPDF} are satisfactory in good agreement with the {\tt MMHT14} model. Perhaps the most surprising discrepancy between our results and the {\tt BBG} model is in the region of $x \approx 0.1$. In spite of this difference, there is quite good agreement for $10^{-4}<x<10^{-2}$, where there is little constraint from data.
There is a strong relationship between the input PDFs parametrization and the uncertainties which will
be obtained. As we mentioned, the valence-quark PDFs errors are
typically computed using the standard error analysis such as Hessian methods. For the present
analysis, we adopted the well-known Hessian method to study the uncertainties of PDFs and
for error calculation proposed by Pumplin, Stump, Tung et al. (PST)~\cite{Pumplin:2001ct,Martin:2003sk}.
In order to have a detailed comparison, we also plotted the obtained error bands for the mentioned groups.
Compared to other results, one finds that the uncertainty band for
both our NLO and NNLO valence parton distributions have become slightly narrower than others, except for {\tt BBG}.
Some groups such as {\tt NNPDF} propose an alternative approach in their analysis for the PDFs
uncertainties, based on an iterative Monte Carlo fitting technique that allows a more robust extraction of PDFs with statistically rigorous uncertainties.
What makes NNPDF differs from others is using neural networks instead of traditional parametrizations.

\begin{table}[htb]
	\begin{tabular}{|c|c|c|}
		\hline
		Parameters &     NLO     & NNLO         \\    \hline  \hline
		$\alpha_{u_v}$ & $ 0.1308 \pm 1.481 \times 10^{-3} $ & $0.1294 \pm 2.657  \times 10^{-3}$  \\
		$\beta_{u_v}$ & $3.629 \pm 2.415 \times 10^{-2}$ & $3.641 \pm 3.397 \times 10^{-2}$    \\
		$\gamma_{u_v}$ & $13.157 \pm 1.387 $  & $15.024\pm 2.109$    \\
		$\eta_{u_v}$ & $ 59.971 \pm 1.443 $  & $68.373 \pm 1.351$    \\
		$\beta_{d_v}$ & $0.6805 \pm 0.248$ & $0.796 \pm 0.2272$    \\
		$\alpha_s({\rm Q}_0^2)$ & $0.2834\pm 0.0117$  & $0.3522 \pm 0.00401$    \\ \hline
		$\chi^2/{\rm d.o.f}$ & $611.71/224 = 2.73$ & $589.61/224 = 2.63$        \\ \hline
		\hline
	\end{tabular}
	\caption{  The optimal values of the input valence-quark PDF parameters for the NLO and NNLO analysis at the scale Q$_0^2$ = 4 GeV$^2$. The corresponding results for the strong coupling constant $\alpha_s(Q_0^2)$ are also shown.  \label{table:fitresults} }
\end{table}

The description of the CCFR~\cite{Seligman:1997mc}, NuTeVs~\cite{Tzanov:2005kr} and CHORUS~\cite{Onengut:2005kv} neutrino and antineutrino dimuon data given by the
NLO and NNLO analyses are shown in Figs.~\ref{fig:xf3ccfr}, \ref{fig:xf3NuTev2006} and \ref{fig:xf3chorus2006}, respectively. Clearly, one can find that the quality of the fit is very good. At moderate to high--$x$, these results are in good consistency with CCFR, NuTeV and CHORUS data over the full energy range, both for the NLO and NNLO analyses.

Another interesting problem is related to the extraction of the value of the Gross-Llewellyn Smith (GLS) sum rule.
The (GLS) sum rule is one of the important characteristics of the deep inelastic neutrino-nucleon scattering. In quark parton model,
the GLS sum rule which is associated with $x F_3$ structure function is given by~\cite{Gross:1969jf}
\begin{equation}\label{eq:GLSSL}
\text{ GLS} (Q^2) = \frac{1}{2} \int_0^1 \frac{xF_3^{\bar{\nu} p + \nu p}(x,Q^2)}{x} dx \,.
\end{equation}
In the work of Ref.~\cite{Leung:1992yx}, authors reported the following result for the
measurement of the GLS sum rule at the scale $|Q^2| = 3$ GeV$^2$,
\begin{equation} \label{eq:GLSCCFR}
	\text{GLS} \, ( |Q^2| = 3 \; {\rm GeV}^2) = 2.5 \pm 0.018 \, (\text{stat.}) \pm 0.078 \, (\text{syst.}).
\end{equation}
The value of GLS sum rule at the scale $|Q^2| =8$ GeV$^2$ was also reported in Ref.~\cite{Londergan:2010cd} as $2.62 \pm 0.15$.
In our work, we obtain $\text{GLS} \, (|Q^2| = 8 \; {\rm GeV}^2) )= 2.64$ for the NLO analysis and $\text{GLS} \, (|Q^2| = 3 \; {\rm GeV}^2) = 2.61$ for the NNLO one, which are in good agreement with the results obtained by other groups.

In conclusion, we would like to stress again, that using the Laplace transform technique and Jacobi polynomial approach, we have shown that these methods  work well in the analysis of the most precise up to now experimental data of the CCFR, NuTeV and CHORUS collaborations for the nonsinglet structure functions of the neutrino-nucleon DIS. The obtained results for valence-quarks densities extracted from the fit of the data turn out to be in good agreement with those from the literature.

%
%
\section{Summary and conclusions}\label{Conclusions}
The main new ingredient of the present analysis comes from the recent results obtained for analytical calculations of the NNLO correction using the Laplace transform technique.
The extracted results for the DGLAP evolution equations in Laplace space were used as input to obtain the $x$ and Q$^2$ evolution of the $xF_3(x, Q^2)$ structure
function. The Jacobi polynomials approach, as an efficiently mathematical tool, is also used to facilitate the analysis and to obtain the Q$^2$ evolution of the $xF_3(x, Q^2)$-function.
We have also utilized the solutions of the DGLAP equations to determine the well-known GLS sum rule with higher-order QCD corrections up to NNLO.
Our theoretical results for the structure function $xF_3(x, Q^2)$ are in good consistency with the neutrino scattering data from the CCFR and NuTeV experiments at Fermilab, and neutrino oscillation search reported by the CHORUS collaboration at CERN.
Although, there are various numerical methods to solve the DGLAP evolution equations to obtain quark and gluon
structure functions, in this manuscript we have shown that the methods of the Laplace transforms technique and Jacobi polynomials approach are also the reliable and alternative schemes to solve these equations, analytically. The advantage of using such a technique is that it enables us to achieve strictly analytical solutions for the PDFs
in terms of the Bjorken-$x$ variable.
Following the methods we used in this paper, the Laplace transformation and Jacobi polynomial approach, we indicated
that these methods work well in which we were able to
extract the valence-quarks distribution functions form the global QCD analysis of neutrino-nucleon scattering data.
We also showed that the obtained results from the present analysis are in good agreement with those from the literature.

%
%
\section*{Acknowledgments}

The authors are especially grateful to Loyal Durand and 
Andrei Kataev for carefully reading the manuscript and fruitful 
discussions. The authors are thankful to the School of Particles 
and Accelerators, Institute for Research in Fundamental 
Sciences (IPM) for financial support of this project.
Hamzeh Khanpour also gratefully acknowledges the University of Science and Technology of Mazandaran for financial support 
provided for this research, and is grateful for the hospitality of the Theory Division at CERN where this 
work has been completed.


%
%
\section*{Appendix A}\label{AppendixA}
Here, We present the Laplace transforms of the NLO and NNLO splitting functions for nonsinglet sectors, denoted by $\Phi_{qq}^{\text{NLO}}$\cite{Curci:1980uw}  and $\Phi _{\text{qq}}^{\text{NNLO}}$\cite{Moch:2004pa}, which we used in (\ref{eq:nonsinglet-laplace-space}). We fixed the usual quadratic Casimir operators to their exact values, using $C_A = 3$, $T_F = f$ and $C_F = 4/3$.  The Laplace transforms of the NLO and NNLO Wilson coefficients  $C_3^{(1)}$\cite{Moch:1999eb} and $C_3^{(2)}$\cite{vanNeerven:1999ca}, are given in (\ref{f1}) and (\ref{f2})respectively. These coefficient functions are used
in (\ref{eq:xf3-laplace-space}).

\begin{widetext}
	
	\begin{eqnarray}
		\Phi_{qq}^{NLO}&=&C_F T_F\Bigg[-\frac{2}{3 (1+s)^2}-\frac{2}{9 (1+s)}-\frac{2}{3 (2+s)^2}+\frac{22}{9 (2+s)}+\frac{20 \left(\gamma _E+\psi (s+1)\right)}{9}+\frac{4}{3} \psi '(s+1)\Bigg]+\nonumber\nonumber\\
		&&C_F{}^2 \Bigg[-\frac{1}{(1+s)^3}-\frac{5}{1+s}-\frac{1}{(2+s)^3}+\frac{2}{(2+s)^2}+\frac{5}{2+s}+\nonumber\\
		&&\frac{2 \left(\gamma _E+\frac{1}{1+s}+\psi (s+1)-(1+s) \psi '(s+2)\right)}{(1+s)^2}+
		\frac{2 \left(\gamma _E+\frac{1}{2+s}+\psi (s+2)-(2+s) \psi '(s+3)\right)}{(2+s)^2}-\nonumber\\
		&&4 \left(\left(\gamma _E+\psi (s+1)\right) \psi '(s+1)-\frac{1}{2} \psi ''(s+1)\right)+3 \psi '(s+1)\Bigg]+\nonumber\\
		&&C_A C_F \Bigg[-\frac{1}{(1+s)^3}+\frac{5}{6 (1+s)^2}+\frac{53}{18 (1+s)}+\frac{\pi ^2}{6 (1+s)}-\frac{1}{(2+s)^3}+\nonumber\\
		&&\frac{5}{6 (2+s)^2}-\frac{187}{18 (2+s)}+\frac{\pi ^2}{6 (2+s)}-\frac{67 \left(\gamma _E+\psi (s+1)\right)}{9}+\nonumber\\
		&&\frac{1}{3} \pi ^2 \left(\gamma _E+\psi (s+1)\right)-\frac{11}{3} \psi '(s+1)-\psi ''(s+1)\Bigg],
	\end{eqnarray}

\begin{eqnarray}
\Phi _{qq}^{NNLO}&=&1295.384\, +\frac{1024}{27 (1+s)^5}-\frac{1600}{9 (1+s)^4}+\frac{589.8}{(1+s)^3}-\frac{1258}{(1+s)^2}+\frac{1641.1}{1+s}-\nonumber\\
&&\frac{3135}{2+s}+\frac{243.6}{3+s}-\frac{522.1}{4+s}+1174.898 \big[\left(\gamma _E+\psi (s)\right)-\left(\gamma _E+\psi (s+1)\right)\big]-\nonumber\\
&&\frac{714.1 \left(\gamma _E+\psi (s+2)\right)}{1+s}+\frac{563.9}{(1+s)^2} \bigg[\gamma _E+\frac{1}{1+s}+\psi (s+1)-(1+s) \psi '(s+2)\bigg]+\nonumber\\
&&f \Bigg[173.927\, +\frac{128}{9 (1+s)^4}-\frac{5216}{81 (1+s)^3}+\frac{152.6}{(1+s)^2}-\frac{197}{1+s}+\frac{8.982}{(2+s)^4}+\nonumber\\
&&\frac{381.1}{2+s}+\frac{72.94}{3+s}+\frac{44.79}{4+s}-183.187 \big[\left(\gamma _E+\psi (s)\right)-\left(\gamma _E+\psi (s+1)\right)\big]+\nonumber\\
&&\frac{5120 \left(\gamma _E+\psi (s+2)\right)}{81 (1+s)}-\frac{56.66}{(1+s)^2} \left(\gamma _E+\frac{1}{1+s}+\psi (s+1)-(1+s) \psi '(s+2)\right)\Bigg]-\nonumber\\
&&\frac{256.8}{(1+s)^4} \bigg[3+2 \gamma _E (1+s)+2 \gamma _E (1+s)^2 \psi (s+1)-(1+s) \big[-1+2 \gamma _E (1+s)\big] \psi (s+1)\nonumber\\
&&+(1+s)^3 \psi (s+1)^2 \psi (s+2)-2 (1+s)^3 \psi (s+1) \psi (s+2)^2+(1+s)^3 \psi (s+2)^3-2 (1+s)^2 \psi '(s+1)+\nonumber\\
&&(1+s)^3 \psi''(s+2)\bigg]+f^2 \Bigg[-\frac{64}{81} \big[\left(\gamma _E+\psi (s)\right)-\left(\gamma _E+\psi (s+1)\right)\big]+\nonumber\\
&&\frac{64}{81} \left(-\frac{51}{16}+\frac{5 \pi ^2}{6}+\frac{3}{2 (1+s)^3}-\frac{11}{2 (1+s)^2}+\frac{7}{1+s}-\frac{3}{2 (2+s)^3}+\right.\nonumber\\
&&\left.\frac{11}{2 (2+s)^2}-\frac{6}{2+s}-3\zeta (3)-5 \psi '(s+2)-\frac{3}{2} \psi''(s+2)\right)\Bigg],
\end{eqnarray}

\begin{eqnarray}\label{f1}
C_3^{(1)}(s)&=&C_F \left(-9-\frac{2 \pi ^2}{3}-\frac{2}{(1+s)^2}+\frac{4}{1+s}-\frac{2}{(2+s)^2}+\frac{2}{2+s}+\right.\nonumber\\
&&3 \left(\gamma _E+\psi (s+1)\right)+\frac{2 \left(\gamma _E+\psi (s+2)\right)}{1+s}+\frac{2 \left(\gamma _E+\psi (s+3)\right)}{2+s}+\nonumber\\
&&\left.\frac{1}{3} \big[\pi ^2+6 \big(\gamma _E+\psi (s+1)\big){}^2-6 \psi '(s+1)\big]+4 \psi '(s+1)\right),
\end{eqnarray}

\begin{eqnarray}\label{f2}
C_3^{(2)}(s)&=&-338.635-\frac{188.64}{s}+\frac{23.532}{(1+s)^4}-\frac{66.62}{(1+s)^3}+\frac{67.6}{(1+s)^2}-\frac{206.1}{1+s}-\frac{576.8}{2+s}-\nonumber\\
&&\frac{31.105 \left(\gamma _E+\psi (s+1)\right)}{s}+\frac{409.6 \left(\gamma _E+\psi (s+2)\right)}{1+s}+\nonumber\\
&&\frac{10.2222}{s} \big[\pi ^2+6 \left(\gamma _E+\psi (s+1)\right){}^2-6 \psi '(s+1)\big]+\frac{94.61}{6+6 s} \big[\pi ^2+6\left(\gamma _E+\psi (s+2)\right){}^2-6 \psi '(s+2)\big]+\nonumber\\
&&\frac{24.65}{(1+s)^2} \bigg[6 \gamma _E{}^2+\pi ^2+\frac{12 \gamma _E}{1+s}+12 \gamma _E \psi (s+1)-6 (1+s) \psi (s+1) \psi (s+2)^2+\nonumber\\
&&6 (1+s) \psi (s+2)^3-6 \left(3+2 \gamma _E (1+s)\right) \psi '(s+2)-12 (1+s) \psi (s+1) \psi '(s+2)+6 (1+s) \psi '' (s+2)\bigg]+\nonumber\\
&&\frac{7.1111}{s} \bigg[2 \gamma _E{}^3+\gamma _E \pi ^2+6 \gamma _E \psi (s+1)^2+2 \psi (s+1)^3+\psi (s+1) \left(6 \gamma _E{}^2+\pi
	^2-6 \psi '(s+1)\right)-\nonumber\allowdisplaybreaks[1]\\
&&6 \gamma _E \psi '(s+1)+2 \psi'' (s+1)+4 \zeta (3)\bigg]+\nonumber\\
&&\frac{7.6}{1+s} \bigg[2 \gamma _E{}^3+\gamma _E \pi ^2+6 \gamma _E \psi (s+2)^2+2 \psi (s+2)^3+\psi (s+2) \left(6 \gamma _E{}^2+\pi
	^2-6 \psi '(s+2)\right)-\nonumber\\
&&6 \gamma _E \psi '(s+2)+2 \psi'' (s+2)+4 \zeta (3)\bigg]+\nonumber\\
&&f \Bigg[46.857\, -\frac{6.3489}{s}+\frac{4.414}{(1+s)^3}-\frac{8.683}{(1+s)^2}-\frac{6.337}{1+s}-\frac{14.97}{2+s}-\frac{8.5926
		\left(\gamma _E+\psi (s+1)\right)}{s}-\nonumber\\
&&\frac{25 \left(\gamma _E+\psi (s+2)\right)}{1+s}-\frac{0.29629}{s} \big[\pi ^2+6 \left(\gamma _E+\psi (s+1)\right){}^2-6 \psi '(s+1)\big]-\nonumber\\
&&\frac{0.808}{6+6 s} \big[\pi ^2+6 \left(\gamma _E+\psi (s+2)\right){}^2-6 \psi '(s+2)\big]+
\frac{9.684}{(1+s)^2} \big[\gamma _E+\frac{1}{1+s}+\psi (s+1)-(1+s) \psi '(s+2)\big]-\nonumber\\
&&\frac{0.021}{1+s} \bigg[2 \gamma _E{}^3+\gamma _E \pi ^2+6 \gamma _E \psi (s+2)^2+2 \psi (s+2)^3+\psi (s+2) \left(6 \gamma _E{}^2+\pi
	^2-6 \psi '(s+2)\right)-\nonumber\\
&&6 \gamma _E \psi '(s+2)+2 \psi'' (s+2)+4 \zeta (3)\bigg]\Bigg].
\end{eqnarray}
\end{widetext}


%
%

%

\end{document}